\documentstyle[twocolumn,prl,aps,floats]{revtex}
\include{epsf}
\begin{document}
\def\la{\langle }
\def\ra{ \rangle }
 \def\bea{\begin{eqnarray}}
\def\eea{\end{eqnarray}}
\def\beq{\begin{equation}}
\def\eeq{\end{equation}}
\def\gmmu{\gamma_{\mu}}
 \def\gmf{\gamma _{5}}
\renewcommand{\topfraction}{0.99}
\renewcommand{\bottomfraction}{0.99}
\twocolumn[\hsize\textwidth\columnwidth\hsize\csname 
@twocolumnfalse\endcsname
  
\title
{\Large Can  Induced $\Theta$  Vacua be Created in Heavy
 Ion Collisions?}
       
\author{K. Buckley, T. Fugleberg and 
A. Zhitnitsky}

\address{~\\Department of Physics and Astronomy, University of 
British Columbia,\\
Vancouver, BC, V6T 1Z1, CANADA} 
\maketitle
\begin{abstract}
\noindent 
The development of the early Universe is a remarkable laboratory for
the study of most nontrivial properties of particle physics.  What is
more remarkable is the fact that these phenomena at the QCD scale can
be, in principle, experimentally tested in heavy ion collisions.  We
expect that, in general, an arbitrary induced $\theta$ vacuum state
($|\theta^{ind}\ra$) would be created in heavy ion collisions, similar
to the creation of the disoriented chiral condensate with an arbitrary
isospin direction.  It should be a large domain with a wrong
$\theta^{ind}\neq 0$ orientation which will mimic the physics of the
early universe immediately following the QCD phase transition when it
is believed that the fundamental parameter, $\theta^{fund}\neq 0$.  We
test this idea numerically in a simple model where we study the
evolution of the phases of the chiral condensates in QCD with two
quark flavors with non-zero $\theta^{ind}$-parameter.  We see the
formation of a non-zero $\theta^{ind}$-vacuum with the formation time
of the order of $10^{-23}$ seconds.
  
\end{abstract} 
\vskip 2mm] 

{\bf 1.}  The realization of the Relativistic Heavy Ion Collider
(RHIC) at Brookhaven opens up some exciting doors in fundamental
physics research.  The possibility of producing and studying
quark-gluon plasma is the most well publicized research area.  But it
is certainly not the only new field that RHIC makes possible.  This
letter is concerned with the possible creation of non-trivial
$\theta^{ind}$-vacua inside the high temperature fireball created at
RHIC.

The idea is very similar to the creation of the
Disoriented Chiral Condensate (DCC) in heavy ion collisons 
\cite{DCC},\cite{RW},\cite{Rajagopal} (see also a 
nice review \cite{Boyanovsky} for a discussion of DCC as an example of
an out of equilibrium phase transition).  DCC refers to regions of
space (interior) in which the chiral condensate points in a different
direction from that of the ground state (exterior), and separated from
the latter by a hot shell of debris.  As we shall see in a moment, for
both cases (DCC and $|\theta^{ind}\ra$-state) the difference in energy
between a created state and the lowest energy state is proportional to
the small parameter $m_q$ and negligible at high
temperature. Therefore, energetically an arbitrary $|\theta^{ind}\ra$
can be formed which is the crucial point for what follows.

Before going into details, we would like to recall some general
properties of the DCC, which (hopefully) can be produced and seen at
RHIC, with an emphasis on the analogy between the DCC and a misaligned
$|\theta\ra$-state.  If the cooling process is very rapid and,
therefore, the system is out of equilibrium, there will be a large
size of the correlated region in which the vacuum condensate
orientation mismatches its zero temperature value.  The absolute value
of the chiral condensate right after the phase transition is expected
to be close to its final (zero temperature) magnitude. However the
vacuum direction of the formed condensate is still misaligned since it
takes a longer time for the vacuum orientation to relax due to the
small free energy difference $\sim m_q$ between the formed and true
vacuum states.


Due to the fact that the absolute value of the chiral condensate right
after the phase transition is expected to be close to the zero
temperature magnitude we can parametrize Goldstone fields by matrix $
U $ in the following way: 
$$ U=e^{i\phi(\vec{n}\vec{\tau})} \; , U
U^{+} = 1 \; , \la \bar{\Psi}_{L}^{i}
\Psi_{R}^{j} \ra 
=  - | \la \bar{\Psi}_{L} \Psi_{R} \ra | \, U_{ij}
$$
The energy density of the DCC is determined by the mass term:
\beq
\label{2}
 E_{\phi}= -\frac{1}{2} Tr( M U + M ^{\dagger}U^{\dagger}) = 
 - 2m  | \la \bar{\Psi}  \Psi  \ra | \cos(\phi)
\eeq
where we put $m_u=m_d=m$ for simplicity.  Eq.(\ref{2}) implies that
any $\phi\neq 0 (mod ~2\pi)$ is not a stable vacuum state because
$\frac{\partial E_{\phi}}{\partial\phi}|_{\phi\neq 0}\neq 0 $,
i.e. the vacuum is misaligned.  On the other hand, the energy
difference between the misaligned state and true vacuum ($\phi=0$) is
small and proportional to $m_q$. Therefore, the probability to create
a state with an arbitrary $\phi$ at high temperature $ T \sim T_c $ is
proportional to $ \exp[- V( E_{\phi} - E_0)/T ] $, where V is 3D
volume, and depends on $\phi$ only very weakly, i.e. $ \phi $ is a
quasi-flat direction. Right after the phase transition when $\la
\bar{\Psi} \Psi \ra $ becomes non-zero, the pion field begins to roll
toward $\phi=0$, and of course overshoots $\phi=0$.  Thereafter,
$\phi$ oscillates. One should expect the coherent oscillations of the
$\pi$ meson field which would correspond to a zero-momentum condensate
of pions. This is exactly what was found in\cite{RW}.  Eventually
these classical oscillations produce real $\pi$ mesons which hopefully
can be observed.

Now we turn to our main point when the $U(1)_A$ phase of the
disoriented chiral condensate is also non-zero and, therefore, the
$|\theta\ra$-vacuum state could be formed\footnote{From now on we omit
the label "ind" for the induced $\theta$.  We hope $\theta^{ind}$ will
not be confused with $\theta^{fund}$ which is zero in our world and
which can not be changed. The simplest way to visualize $\theta^{ind}$
is to assume that right after the QCD phase transition the flavor
singlet phase of the chiral condensate is non-zero in a
macroscopically large domain. This phase is identified with
$\theta^{ind}$. This identification is a direct consequence of the
transformation properties of the fundamental QCD lagrangian under
$U(1)_A$ rotations by which the chiral singlet phase can be rotated
away at the cost of introducing $\theta^{ind}$.}.  The production of
non-trivial $\theta$-vacua would occur in much the same way as
discussed above.  The new element is that in addition to chiral fields
differing from their true vacuum values the $\theta$-parameter of QCD,
which is zero in the real world, becomes effectively nonvanishing in
the macroscopically large region.

In this letter we show that in a simplified numerical model
non-trivial $\theta$-vacua can be realized.  While this is a
simplified model which differs somewhat from the real physics
associated with high energy ion collisions, there is no reason to
believe that the non-trivial $\theta$-vacua cannot be realized at
RHIC.  Our results also give a very rough estimate of the time it
takes for these non-trivial $\theta$-vacua to be formed.  In order to
have a hope of observing them they must form within the time that the
central region of the fireball is isolated from the true vacuum.

{\bf 2.}  To take into account the $U(1)_A$ phase associated with
$|\theta\ra $-vacua we choose the matrix $ U_{ij} $ in the form
$U=diag \; (e^{i\phi_i} ) $ with $\sum_{i=1}^{N_f}\phi_i$ in general
non-zero.  The energy density of the misaligned vacuum is determined
in this case by the following low-energy
potential\cite{QCD},\cite{QCDanalysis}:
\bea
\label{potential}
V(\phi_i,\theta)\!&=&\!\! -E\cos\! \left[ \frac{1}{N_c}\! \left(\sum_{i=1}^{N_f} \phi_{i}\! -\! 
\theta \right)
\! +\! \frac{2 \pi}{N_c} \, l
  \right] \!\!
-\!\! \sum_{i=1}^{N_f} \!M_{i} \cos \phi_i \nonumber\\
 &&   l= 0,1, \ldots , N_c-1,
\eea
where $E=\langle b \alpha_s/(32\pi)  G^2 \rangle\sim 10^{-2}GeV^4$ is
much larger than  $M_i=- m_q \langle \bar{\Psi}\Psi \rangle
\sim 10^{-3}GeV^4$. Here $b=11/3 N_c-2/3 N_f,\,m_q\sim 5 MeV,\,
\la \bar{\Psi}\Psi \ra\simeq -(240 MeV)^3$. 
The crucial point is that the $\theta$ parameter appears only in the
combination, $\sum \phi_{i} - \theta$.  This is a direct consequence
of the transformation properties of the chiral fields under $U(1)_A$
rotations.  To convince the reader that (\ref{potential}) does indeed
represent the anomalous effective low energy Lagrangian, three of its
most salient features are listed below:
\\[0mm]
\indent i)
Eq. (\ref{potential}) correctly reproduces the Witten-Di
Vecchia-Veneziano effective chiral Lagrangian
\cite{Wit2} in the large $ N_c $ limit; \\[0mm]
\indent ii)
it reproduces the anomalous conformal and chiral Ward identities of
$QCD$;\\[0mm]
\indent iii) 
it reproduces the known dependence in $\theta$ ({\it ie.} $2\pi$
periodicity of observables) 
\cite{Wit2}.  
As mentioned above, in the large $N_c$ limit the effective Lagrangian
(\ref{potential}) takes the Witten-Di Vecchia-Veneziano quadratic
form.  We would like to remark here that the exact form is not
important for our present purposes as long as the $\theta$-parameter
appears in combination $\sum \phi_{i} - \theta$ and the parameter
$E\gg M_i$.
 
The most important difference between Eqs. (\ref{2}) and
(\ref{potential}) is the presence of the parametrically large term $
\sim E\gg m_q | \la \bar{\Psi} \Psi \ra |$ in the expression for
energy (\ref{potential}), describing the $U(1)_A$ phase of the
disoriented chiral condensate.  This term does not go away in the
chiral limit and provides a non-zero mass for the $\eta'$ meson which
is expressed in terms of the parameter, E.  It was for exactly this
reason that it was thought until recently\cite{HZ} that the
non-trivial $\theta$-vacua would involve too large an energy cost to
be produced because of the large parameter associated with the
$\theta$ parameter.
 
The key point is the following. For arbitrary phases $ \phi_i$ the
energy of a misaligned state differs by a huge amount $\sim E $ from
the vacuum energy.  Therefore, apparently there are no quasi-flat
($\sim m_q$) directions along $\phi_i$ coordinates, which would lead
to the long wavelength oscillations with production of a large size
domain. However, when the relevant combination $(\sum_i\phi_i-\theta)$
from Eq.(\ref{potential}) is close by an amount $ \sim {\cal O}(m_q) $
to its vacuum value, a Boltzmann suppression due to the term $ \sim E
$ is absent, and an arbitrary misaligned $|\theta\ra$-state can be
formed.

Indeed, in the limit $M_i \ll E$, because of the large parameter in
the first term the vacuum state that is favored for non-zero $\theta$
is the solution $\bar{\phi_i}\approx\theta/N_f$.  Substituting this
back in to the potential we reproduce the well-known result\cite{Wit2}
for the vacuum energy as a function of $\theta$.
\beq
\label{3}
V(\theta)\approx -E 
- \sum M_{i} \cos(\theta/N_f) 
\eeq
This shows that the energy cost of creating a non-trivial
$\theta$-vacuum goes like the much smaller parameter $M_i$ in
agreement with a general theorem that the $\theta$ dependence appears
only in combination with $m_q$ and goes away in the chiral limit.
 
At this point we can apply the same philosophy as for DCC.  The chiral
fields\footnote{If $\theta\neq 0$, the Goldstone fields are not
exactly the pseudoscalar fields, but rather are mixed with the
scalars; the mixing angle between the singlet and octet combinations
also depends on $\theta$, see \cite{QCDanalysis} for detail.}
$\phi_i$ are allowed to take random values and after the phase
transition begin to roll toward the true solution
$\bar{\phi_i}\approx\theta/N_f$ and of course overshoot it. The
situation is very similar to what was described for the DCC with the
only difference that in general we expect an arbitrary
$|\theta\ra$-disoriented state to be created in heavy ion collisions,
not necessarily the $|\theta=0\ra$ state. The difference in energy
between these states is proportional to $m_q$ from (\ref{3}).

To be more quantitative, one should study the evolution of the chiral
fields with time.
For this paper we assume the following situation.
The rapid expansion of the high energy shell leaves behind an
effectively zero temperature region in the interior which is isolated
from the true vacuum.  The high temperature non-equilibrium evolution
is very suddenly stopped, or ``quenched'', leaving the interior region
in a non-equilibrium initial state that then begins to evolve
according to (almost) zero temperature Lagrangian dynamics.  Starting
from an initial non-equilibrium state we can study the behavior of the
chiral fields using the zero temperature equations of motion.  The
equations of motions are non-linear and cannot be solved analytically
but we can solve them numerically in order to determine the behavior
of the fields.  The production of a non-trivial $\theta$-vacuum is
indicated by the fact that the chiral fields relax to constant and equal
non-zero values on a time scale over which spatial oscillations of the
fields vanish.  This fact confirms the self consistency of
our approach.  First we assumed that the singlet phase of the chiral
condensate is non-zero on a macroscopically large domain. 
We then calculated the evolution of the chiral phases in this background.
Finally we closed the circle by realizing that the chiral fields relax
to constant, equal, non-zero values as mentioned above.
The formation of a non-perturbative condensate is also supported
by observation of the phenomenon of coarsening (see below) and
by a test of volume-independence of our results. 

{\bf 3.}  The equations of motion for the phases of the chiral
condensate with two quark flavors consists of two coupled second order
nonlinear partial differential equations:
\beq
\ddot{\phi}_i - \nabla^2 {\phi_i} + \gamma \dot{\phi_i} + \frac{d}{d\phi_i} 
V(\phi_j,\theta)=0 \; \; \; \; \; i=1,2
\eeq
where $\nabla^2$ is a three dimensional spatial derivative and the
potential is given in (\ref{potential}).  Emission of pions and
expansion of the domain will contribute to the damping, $\gamma$, as
might other processes.  We do not know exactly how they would
contribute but we simulate these unknown effects by including a
damping term with a reasonable value for the damping constant, $\gamma
\sim\Lambda_{QCD}\sim 200 ~ MeV$.

The initial data for each of the chiral fields $\phi_i$ is chosen on a
3D grid of $16^3$ points.  The initial data consisted of random values
of $\phi_i$ and $\dot{\phi_i}=0$.  The initial data was evolved in
time steps using a Two-Step Adams-Bashforth-Moulton
Predictor-Corrector method for each grid point with the spatial
laplacian approximated at each grid point using a finite difference
method.  We used periodic spatial boundary conditions.

The grid spacing was determined by the length of a side of the spatial
grid which was varied in order to vary the volume.  The size of the
time step between successive spatial grids was much smaller than the
spatial grid spacing and was fixed at about $10^{-5} MeV^{-1} $.

We evolved the data for 8000 time steps and then applied a Fast
Fourier Transform \cite{fourier} to the spatial data at evenly spaced
time steps.  We then binned the data in small increments of the
magnitude of the wave vector in order to obtain the angular averaged
power spectrum.

This procedure was carried about for different volumes.  In all cases
the results were qualitatively the same.  We saw an initial growth of
long wavelength modes as in \cite{RW} and subsequent damped
oscillation of all modes.  The $|\vec{k}|=0$ modes oscillate and
approach the equilibrium values of the fields.  They exhibit this
behaviour in the same time frame in which the Fourier coefficients of
the modes with non-zero wave vectors fall to a tiny fraction of the
zero mode coefficient.  This qualitative behavior occurs for different
total volumes and grid sizes suggesting that this behavior is not due
to finite size effects.

We should note that our $|\vec{k}|=0$ mode is really only a quasizero
mode as it is obtained in a finite spatial volume with periodic
boundary conditions.  However, our quasizero mode approaches the same
value irrespective of the total spatial volume indicating that this
really is a condensate.  If it were not we would expect the value of
the coefficient to decrease when the volume of the system increases.

\begin{figure}
\epsfysize=1.55in
\epsfbox[ 35 545 348 733]{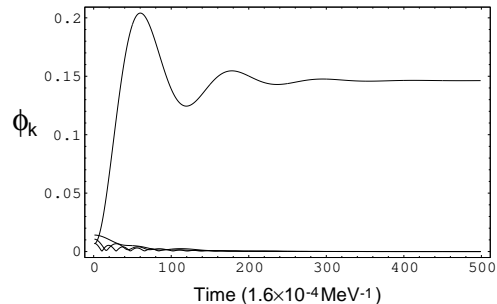}
\caption{$|\phi_k|$ is shown for various  $|\vec{k}|$ as a function of time. 
Notice that the zero mode settles down to $\bar{\phi_i}\approx\theta/N_f$. }
\label{Evolution}
\end{figure}

The evolution of the Fourier modes of the $\phi$ fields is shown in
Fig.\ref{Evolution} for the specific case of $\theta=2\pi/16$ and a
spatial grid of 10 fm on a side.  The initial values of the $\phi$
fields were randomly chosen within the range $-7\pi/16$ to $7\pi/16$.
The zero mode clearly settles down to a non-zero value.  All higher
momentum modes vanish extremely rapidly and are negligible long before
the zero mode settles down to its equilibrium value.

\begin{figure}
\epsfysize=1.55in
\epsfbox[ 15 570 321 759]{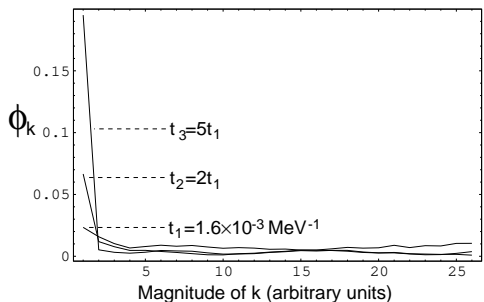}
\caption{Above we demonstrate that the system exhibits the coarsening 
phenomenon. The data was sampled at three times within the first 1000 
timesteps of the evolution.}
\label{Fourier}
\end{figure}

The instantaneous distribution of Fourier modes for the evolution above
 is shown in Fig.\ref{Fourier}
at a few different times.  This graph clearly shows the amplification of
 the zero mode as time increases. This phenomenon of coarsening
and the formation of a nonperturbative condensate
 is
very similar to earlier discussions in ref.\cite{Boyanovsky}.

\begin{figure}
\epsfysize=1.55in
\epsfbox[ 42 543 359 737]{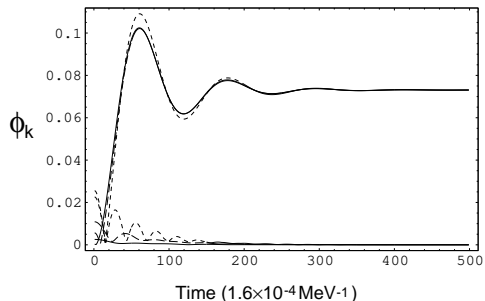}
\caption{The zero mode and a non-zero mode are shown as a function of time for three 
different volumes. The heavily dashed line represents the smallest
volume, the medium dashed line the middle volume, and the solid line
represents the largest volume.\ This graph illustrates the volume
independence of the zero mode.}
\label{Volumes}
\end{figure}

In Fig.\ref{Volumes} we plot $|\phi_k|$ as a function of time for
three different volumes.  We chose $\theta=\pi/16$ and the volumes $(8
fm)^3$, $(16 fm)^3$, and $(32 fm)^3$. For each volume, we plotted the
zero mode and the same non-zero mode. Notice that the zero mode is
independent of the volume of the system, while the magnitude of the
non-zero mode decreases with increasing volume.  This is the signature
that a real condensate has been formed.

For a total volume of $(10 fm)^3$ and $\theta=2\pi/16$ the time for
relaxation from the initial non-equilibrium state following the quench
to the non-trivial $\theta$-vacuum is approximately 0.064
MeV$^{-1}\approx 4\times 10^{-23} s$.  This seems to be of the same
order of magnitude as the time which we might expect the central
region of the fireball to be isolated from the usual vacuum.  The
volume we have used is just at the upper limit of what we would expect
at RHIC.  As well we have not attempted to correctly account for the
collision geometry at RHIC either.  However, our simplified
calculation suggests the possibility of producing non-trivial
$\theta$-vacua.

{\bf 4.}  We have shown through a numerical calculation of the zero
temperature equations of motion with a non-zero induced $\theta$
parameter that the chiral fields $\phi$ after a quench from high
temperature go to a spatially constant non-zero value related to the
$\theta$-parameter.  This occurs on a time scale of the order of
$10^{-23}$ seconds.  The fact that all other non-zero modes fall to
negligible values long before indicates that we have formed a
condensate or a non-trivial $|\theta\ra$ vacuum state.

The most intriguing question is: ``What is the signature of the
produced $\theta$ state?''  First of all, due to the fact that the
$\theta$-vacua are odd under charge conjugation times parity, CP,
their decay must produce some CP odd correlations suggested in
ref.\cite{Pisarski}.  However, one should expect that the signal will
be considerably (if not completely) washed out by the re-scattering of
the pions and their interactions in the final states, which mimic true
CP-odd effects.  In practice it is quite difficult to overcome the
problem of separating a true CP violation from its simulation due to
the final state interactions.  A more promising direction is to
to look at $\eta(\eta')\rightarrow\pi\pi$\cite{Pisarski,arz} decays
which are strongly forbidden in our world, but nevertheless will be
of order one if $\theta\neq0$. Indeed, one can show \cite{arz} that
\bea
 \Gamma(\eta\rightarrow\pi\pi)= 
 \frac{2\left(1-\frac{4m_{\pi}^2(\theta)}{m_{\eta}^2(\theta)}
\right)^{\frac{1}{2}} }{3\pi m_{\eta}(\theta)}\left(\frac{m_q
\sin\frac{\theta}{2} \la 0|\bar{q} q |0 \ra}{f_{\pi}^3}\right)^2
\nonumber \\
\sim 0.5~MeV (\sin\frac{\theta}{2})^2,~~~~~~~~~~~~~~~~~~~~~~~
\label{eta2}
\eea
and therefore the effect could be quite noticeable.  We should remark
here that all masses in the $\theta$ vacuum state are shifted in
comparison with their values at $\theta=0$\cite{QCDanalysis}.  Formula
(\ref{eta2}) reduces to the corresponding expression of
reference\cite{svz} in the limit $\theta\rightarrow 0$.  An analogous
calculation for $\eta'\rightarrow\pi\pi$
leads to a  similar
numerical estimation for $ \Gamma(\eta'\rightarrow\pi\pi) 
\sim 2~MeV (\sin\frac{\theta}{2})^2 $
which is almost an order of magnitude larger than the full width of the
$\eta'$ meson in the $\theta=0$ world.  More importantly, this
width is exclusively due to the CP odd decay
$\Gamma(\eta'\rightarrow\pi\pi)$.


\end{document}